\documentclass[a4paper,11pt]{article}

\usepackage[a4paper, total={6.5in, 10.5in}]{geometry}
\usepackage[T1,T2A]{fontenc}
\usepackage{textcomp}

\usepackage{amsmath}
\usepackage{amssymb}
\usepackage{mathrsfs}  
\usepackage{braket}

\usepackage[affil-it]{authblk}

\usepackage{hyperref}
\usepackage{graphicx}
\usepackage{color}
\usepackage[dvipsnames]{xcolor}

\usepackage{subfigure}
\usepackage{multirow}
\usepackage{todonotes}

\newcommand*{\email}[1]{\href{mailto:#1}{\nolinkurl{#1}} } 

\usepackage{slashed}
\usepackage{wrapfig}
\usepackage{textcomp}
\usepackage{amsmath,amssymb,fontenc,times, mathptmx}
\usepackage{float}
\usepackage{verbatim}
\usepackage{todonotes}

\begin{document}

\title{Problems with  Higgsplosion}

\author[a,b]{Alexander Belyaev\footnote{\email{belyaev1967@gmail.com}}}
\author[c]{Fedor Bezrukov\footnote{\email{Fedor.Bezrukov@manchester.ac.uk}}}
\author[c]{Chris Shepherd\footnote{\email{christopher.shepherd-3@postgrad.manchester.ac.uk}}}
\author[a]{Douglas Ross\footnote{\email{D.A.Ross@soton.ac.uk}}}

\affil[a]{University of Southampton, School of Physics and Astronomy,\\
  Southampton SO17 1BJ, UK}
\affil[b]{Particle Physics Department, Rutherford Appleton Laboratory, Chilton, Didcot, Oxon OX11 0QX, UK}
\affil[c]{The University of Manchester, School of Physics and Astronomy,
  Oxford Road, Manchester M13 9PL, UK}

\maketitle



\begin{abstract}
  A  recent calculation of the multi-Higgs boson production in
  scalar theories with spontaneous symmetry breaking
  has demonstrated the fast growth of the cross section with the Higgs multiplicity
  at sufficiently large energies,  called ``Higgsplosion''.
  It was argued that ``Higgsplosion'' solves the Higgs hierarchy and fine-tuning problems.
  In our paper  we argue that:
  a) the formula for ``Higgsplosion'' has a limited applicability and inconsistent with unitarity
  of the Standard Model;
  b) that the contribution from ``Higgsplosion'' to the imaginary part of the Higgs boson propagator 
  cannot be re-summed  in order to furnish a solution of the Higgs hierarchy and fine-tuning problems.
\end{abstract}

\newpage

\section{The amplitude behaviour  with the large scalar multiplicity}

One of the flaring questions for the modern elementary particle physics is  the question about the  energy scale of  new physics.   All current experiments are in excellent agreement with the Standard Model (SM).  Moreover, the Higgs mass $m_H\simeq 125$\,GeV means that all the couplings of the theory are small above the electroweak scale, and  perturbative calculations in  non-abelian QFT, which
is the core of the SM, should provide a consistent approach.  Most of the coupling constants of the theory become smaller with increasing energy. The only two
 couplings which grow with the energy scale are the $U(1)$ hypercharge coupling constant and the Higgs self coupling $\lambda$.  
However, the scale  of new physics related to this coupling evolution with the energy -- the Landau pole -- is proportional to $\exp(1/\lambda)$ and significantly exceeds the Planck scale.  Therefore, it is normally assumed that SM can be trusted as a perturbative QFT at all energies that can, even hypothetically, be  probed in collisions.  The only scale that may appear in the SM framework is the one  associated with the metastability of the EW vacuum, but this scale, even if present, is very large $\sim10^{10}$\,GeV.

At the same time, it has long been known that theories of self-interacting scalars (which also include  the Higgs boson of the SM) have problems with the  application of perturbation theory at high energies. The first observations of subtleties in the scalar multi-particle production demonstrated that at  the tree level, owing to the large number of contributing diagrams, the $n$-particle
amplitudes have factorial dependence on the number of particles.
\cite{Cornwall:1990hh,Goldberg:1990qk,Voloshin:1992mz,Argyres:1992np,Brown:1992ay}
\begin{equation}
\label{Atree0}
A_{n}^\mathrm{tree}(0) = n!\left(\frac{\lambda}{8}\right)^{\frac{n-1}{2}}
\end{equation}
This factorial growth of the amplitude indicates the  breakdown of the
usual perturbative calculations for $n\gtrsim\lambda^{-1}$.  It was
found \cite{Libanov:1994ug,Libanov:1995gh,Son:1995wz} that the
corresponding $1\to n$ cross-section
 can be written in  exponential form
\begin{equation}
\label{sigma}
\sigma(E,n) \propto \exp\left(\frac{1}{\lambda} F(\lambda n,\epsilon)\right),
\end{equation}
where $\epsilon \equiv(E-nm_H)/nm_H$ is the average kinetic energy of the final-state Higgs particles.
The function $F(\lambda n,\epsilon)$ was obtained by following a specific
semiclassical approach \cite{Son:1995wz} valid in the limit
\begin{equation}  \lambda\to0,\; n\to\infty,
\quad \text{with fixed}\quad
\lambda n,\;
\epsilon.
\end{equation}
Moreover, there is a conjecture \cite{Libanov:1995gh}, that to 
exponential precision the result does not depend on the details of the
initial state, given that the initial number of particles is
small and therefore, without loss of  generality,  one can focus on calculation of $1\to n$
process, even though the initial particle is off-shell.  
For small $\lambda n\ll1$ and small energies of the final particles $\epsilon\ll1$ the exponent of the cross-section is \cite{Libanov:1994ug,Libanov:1995gh,Son:1995wz}\footnote{We quote here
	the result for the theory with spontaneous symmetry breaking, which
	was used  in the recent calculations~\cite{Khoze:2017ifq,Khoze:2018kkz}.}
\begin{equation}
\label{FHG2}
F(\lambda n,\epsilon) =
\lambda n\ln\frac{\lambda n}{16}-\lambda n
+\frac{3}{2}\left(\ln\frac{\epsilon}{3\pi}+1\right)
-\frac{25}{12}\lambda n \epsilon
+2  B\lambda^2n^2
+O(\lambda^3n^3)+O(\lambda^2n^2\epsilon)+O(\lambda n\epsilon^2),
\end{equation}
where
\[
B = \frac{\sqrt{3}}{8\pi}.
\]
As $\lambda n \to0$, $F(\lambda n,\epsilon)\to -\infty$ and  the
cross-section Eq.\eqref{sigma} is exponentially suppressed, 
whilst in the opposite regime for large $\lambda n$ the 
cross section grows exponentially, thereby contradicting the unitarity of the theory,
at least at the level of perturbation theory.

The expression Eq.\eqref{FHG2} for $F(\lambda n,\epsilon)$ is valid for $\lambda n \ll1$, $\epsilon\ll1$.  The logarithmic and lowest order terms correspond to tree level contributions, the term of the order $O(\lambda^2n^2)$ is the first radiative correction.  Note, that in
the range of the  validity of  Eq.\eqref{FHG2} the function  $F(\lambda n,\epsilon)$
 is negative.  At tree level (for $\lambda n\ll 1$) the energy dependence  for arbitrary energies $\epsilon$ was found in
\cite{Bezrukov:1995ta,Bezrukov:1999kb} and again leads to an
exponentially suppressed result.   However, the problem of finding the expression for  arbitrary large $\lambda n$ and $\epsilon$ is still open.

Recently authors of  \cite{Khoze:2017ifq,Khoze:2018kkz}  have extended  the thin-wall approximation of \cite{Gorsky:1993ix}
and have found the cross-section for  the opposite,
$\lambda n\gg 1$ limit:
\begin{equation}
\label{FKhoze}
F(\lambda n,\epsilon) = \lambda n\left(
\ln\frac{\lambda n}{4}+0.85\sqrt{\lambda n}-1
+\frac{3}{2}\left(\ln\frac{\epsilon}{3\pi}+1\right)
-\frac{25}{12}\epsilon
\right).
\end{equation}
An important  feature of this solution is the 
increase of $F(\lambda n,\epsilon)$  at sufficiently  large
$\lambda n$ for a fixed value of $\epsilon$. This result was then used  to argue that at large multiplicities (or, equivalently, large energies 
$E\sim n(\epsilon+m_H)$) the $1\to n$ width grows exponentially. One should  note that the thin-wall semi-classical solution, leading to Eq.\eqref{FKhoze} exists only in the $\lambda\phi^4$ theory with spontaneous symmetry breaking in 3+1 dimensions.

We would like to stress, however, that non-vanishing $\epsilon$ is required for the result of Eq.\eqref{FKhoze} to be  positive, since at  zero $\epsilon$ the logarithmic term is infinitely negative which gives  zero cross-section at the threshold.  At the same time the contribution $0.85\sqrt{\lambda n}$ in Eq.\eqref{FKhoze} was  obtained at the kinematical threshold, that is for $\epsilon\to0$. 
This is a subtle point. One should also note that the full result of Eq.\eqref{FKhoze} is obtained from  a  combination of  the large $\lambda n$ contribution with the tree level result, which has the factorized form
\[
F(\lambda n,\epsilon) = \lambda n( f_0(\lambda n)+f(\epsilon) ).
\]
This form is valid at tree level and at one loop (c.f.\ Eq.\eqref{FHG2}). We would now like to point out that
 higher order quantum corrections are expected to contain terms
which depend {\it both}  on $\epsilon$ and $\lambda n$, e.g.\ terms like $O(\lambda^2n^2\epsilon)$ in Eq.\eqref{FHG2}. Such terms could play an important role. We  argue here that without the knowledge of these terms it is not possible to determine the validity region of the result Eq.\eqref{FKhoze} 
with respect to the value of $\epsilon$.  We discuss this in detail in the next section. Such mixed terms may prevent the exponential growth of the cross-section.
The exponential growth of the $1\to n$ width was suggested to be by itself a solution to the hierarchy problem in \cite{Khoze:2017tjt} where authors conclude that such exponential growth of the self-energy leads,  after resummation, to exponential suppression of the scalar propagators at high energies.

In this paper we review in detail the 
validity and consequences of  such fast-growing amplitudes in the context of unitary, local and Lorentz invariant quantum field theory.

\section{Unitarity and 1PI resummation}

It has been known for many years \cite{Goldberg:1990qk} that exponentially growing amplitudes lead to a violation of unitarity. In \cite{Khoze:2017tjt} the authors have proposed a  mechanism to recover unitarity  through the effect of  the off-shell $1\rightarrow n$ amplitude on the re-summed scalar Feynman propagator. The authors suggested that if the two-point function falls off faster with energy than the amputated $1\rightarrow n$ matrix element, unitarity can be 
restored via the so-called Higgspersion mechanism. 

However, this argument requires a propagator which falls off faster than the amputated $1\rightarrow n$ matrix element. In other words, we require the two-point function to be decaying exponentially with energy. This is a peculiar form of the two-point function that is known to cause problems with unitarity \cite{Stelle1977}. However, it has been proposed \cite{Khoze:2017tjt} that this form appears in a theory with exploding amplitudes.

The problem we see here is the following. An exponentially decreasing propagator has been obtained in \cite{Khoze:2017tjt} because the authors 
have used the perturbation theory to sum up single-particle irreducible (1PI) Green's functions, which is a valid procedure only for a convergent geometric series. 
Namely, it has been claimed that the exact two-point function $\Delta_F (p^2)$ can be obtained from the 1PI Green's function $\Sigma (p^2)$ via
\begin{equation}
\Delta_F (p^2) = \int \mathrm{d}^4 x\,\, e^{i p\cdot x} \left\{ \theta (x_0) \left<0| \phi(x) \phi(0)|0\right> +   \theta (-x_0) \left<0| \phi(0) \phi(x)|0\right>
\right\} = \frac{i}{p^2 - m_0 ^2 - \Sigma (p^2)}
\end{equation}
where $m_0$ is the bare mass of the theory. However, if $\Sigma$ is exponentially growing with $p^2$, at sufficiently large $p^2$ this series is no longer convergent. In this case, one may not use the re-summed form of the above expression. Since resummation 
is not valid, instead of exponentially falling with $p^2$, $\Delta_F$ will uncontrollably grow with $p^2$. This leads to unitarity violation  of the Higgsploding theory,
	assuming that Eq.\eqref{FKhoze} is valid for large $\lambda n$
	values and non-vanishing $\epsilon$.
Under this assumption, one may ask whether the aforementioned problem is related to the application of the  perturbation theory where it is not valid. It is illustrative to examine the functional form of the two-point function using  non-perturbative ``language'' of dispersion relations. In this procedure we closely follow \cite{barton1965introduction}. Consider the momentum-space Feynman propagator
\begin{equation}
\Delta_F (p^2) = \int \mathrm{d}^4 x\,\, e^{i p\cdot x} \left\{ \theta (x_0) \left<0| \phi(x) \phi(0)|0\right> +   \theta (-x_0) \left<0| \phi(0) \phi(x)|0\right>
\right\}
\end{equation}
where we anticipate that the $\Delta_F$ is Lorentz invariant and hence only a function of $p^2$. 

Using the integral representation of the $\theta$-function,
\begin{equation}
e^{ip_0\cdot x_0} \, \theta(\pm x_0) = \frac{1}{2 \pi i} \int \mathrm{d} p_0 ' \, \, e^{i p_0' \cdot x_0} \, \frac{1}{p_0' - p_0  \mp i \epsilon}
\end{equation}
one has
\begin{equation}
\Delta_F (p^2) = \int \mathrm{d}^4 x  \int \frac{\mathrm{d} p_0'}{2 \pi i}\, e^{-i \vec{p}\cdot\vec{x} + i p_0' \cdot x_0}\left\{ \frac{ \left<0| \phi(x) \phi(0)|0\right>}{p_0' - p_0 + i \epsilon} - \frac{ \left<0| \phi(0) \phi(x)|0\right>}{p_0' - p_0 - i \epsilon}
\right\}
\end{equation}
Setting the variable of integration $\vec{x} \rightarrow -\vec{x}$ in the second term and using translation invariance of the vacuum,
\begin{equation}
\Delta_F (p^2)  \ = \  \int \mathrm{d}^4 x \, \int \frac{\mathrm{d} p_0'}{2 \pi i}\,   \left<0| \phi(x) \phi(0)|0\right> \left\{
\frac{e^{i p_0' \cdot x_0-i \vec{p}.\vec{x}}}{p_0' - p_0 + i \epsilon} - \frac{e^{i p_0' \cdot x_0+i \vec{p}.\vec{x}}}{p_0' - p_0 - i \epsilon}
\right\}
\end{equation}
Now we insert   a complete set of (\textit{in} or \textit{out}) states. In the language of \cite{Khoze:2017tjt}, this corresponds to a kinematically-unique one-particle state, plus a continuum of multi-particle states. We let $\sigma_n$ denote all the internal quantum numbers of an n-particle state, including its phase space. Assuming that $ \left|\left<0| \phi(0)|n,\sigma_n\right>\right|^2$ is Lorentz invariant,
\begin{equation}
\Delta_F (p^2) = \int \mathrm{d}^4 x\,  \int \frac{\mathrm{d} p_0'}{2 \pi i}\,\, \sum_{n,\sigma} \left|\left<0| \phi(0)|n,\sigma_n\right>\right|^2 (p_n ^2) e^{-i p_n \cdot x}
\left\{
\frac{e^{i p_0' \cdot x_0-i \vec{p}\cdot\vec{x}}}{p_0' - p_0 + i \epsilon} - \frac{e^{i p_0' \cdot x_0+i \vec{p}\cdot\vec{x}}}{p_0' - p_0 - i \epsilon}
\right\}
\end{equation}
where $p_n$ is the total four-momentum of the n-particle state. 

In the case of \cite{Khoze:2017tjt}, complications will arise due to the divergence of this integrand. To see how difficulties appear, let us consider the scenario where $\sum_{n,\sigma}  \left|\left<0| \phi(0)|n,\sigma_n\right>\right|^2$ is a polynomial of order $N$ in $p^2$.

Exchanging $p_0' \rightarrow - p_0'$ and $x_\mu \rightarrow -x_\mu$ in the second term gives
\begin{eqnarray}
\Delta_F (p^2) & = & \int \mathrm{d}^4 x \,  \int \frac{\mathrm{d} p_0'}{2 \pi i}\,\sum_{n,\sigma}
e^{i ( p_0'- p_{n,0})\cdot x_0 -i (\vec{p}- \vec{p_n})\cdot \vec{x}}
\left|\left<0| \phi(0)|n,\sigma_n\right>\right|^2 (p_n ^2)
 \nonumber  \\  & & \ \times
\left\{
\frac{1}{p_0' - p_0 + i \epsilon} + \frac{1}{p_0' + p_0 + i \epsilon}
\right\}   \label{divForm}
\end{eqnarray}
Combining both terms in the curly bracket and making the $i\epsilon$ prescription implicit,
\begin{equation} \label{divFormTwo}
\Delta_F (p^2)  =  \int \mathrm{d}^4 x \, \int \frac{\mathrm{d} p_0{'}^2}{2 \pi i}\,\sum_{n,\sigma}
e^{i ( p_0'- p_{n,0})\cdot x_0 -i (\vec{p}- \vec{p_n})\cdot \vec{x}}
\left|\left<0| \phi(0)|n,\sigma_n\right>\right|^2 (p_n ^2) 
\,\frac{1}{\left(p_0'^2 - p_0^2\right)}
\end{equation}
At this point, one might be tempted to swap the order of integration and perform the $x$-integral. However, the remaining integrand would be an order $N-1$ polynomial in $p'^2$. This integrand is not convergent at $p_0' = \pm \infty$, so the straightforward change of the integration order is not valid here. Before we can swap the order of integration, we must perform $N$ subtractions of the form
\begin{equation} \label{subtractions}
\frac{1}{p_0'^2 - p_0^2} = \frac{1}{p_0'^2} + \frac{p_0^2}{p_0'^2 (p_0'^2 - p_0 ^2)} 
\end{equation}
In this way, Eq.\eqref{divFormTwo} may be written as $(p_0^2)^N$ times a convergent integral, plus an order $N-1$ polynomial in $p_0^2$ where the coefficients are functions of $\Delta_F(0)$. For example, the first term in Eq.\eqref{subtractions} simply gives $\Delta_F (0)$.

The contribution from the convergent integral is
\begin{equation}
(p_0 ^2)^N 	\int \mathrm{d}p_0{'}^2 \left\{(2 \pi)^3 \sum_{n,\sigma} \left|\left<0| \phi(0)|n,\sigma_n\right>\right|^2 (p{'} ^2)\,\, \delta^{(4)} (p' - p_n)
\right\} \times \frac{ -i }{(p_0{'}^2 - p_0^2)(p_0'^2)^N}
\end{equation}
where $p{'}^\mu\equiv(p_0', \vec{p})$ and we recognize the term in the curly brackets as the Kallen-Lehmann spectral function $\rho (p'^2)$. Given that $\vec{p}$ is fixed, one may change the variable of integration from $p_0'^2$ to $p'^2$, giving 
\begin{multline} \label{subForm}
\Delta_F (p^2) = \Delta_F (0) + p^2 \Delta_F ^{(1)} (0) +   (p^2)^2 \Delta_F ^{(2)}(0) + .... 
+ (p^2)^N \int \mathrm{d} p'^2 \,\,\rho(p'^2) \frac{-i}{(p'^2)^N (p'^2 - p ^2)} 
\end{multline}
From this form it is evident that if $\rho(p^2)$ is an order-$N$ polynomial in $p^2$, knowledge of $\rho(p^2)$ only defines the two-point function up to some order-$N$ polynomial. The functional form of $\Delta_F (p^2)$ is allowed to change dramatically without any change in the amputated $1\rightarrow n$ matrix element. 

In the case of \cite{Khoze:2017tjt}, the situation is even more extreme. In this case, the spectral function $\rho(p^2)$ is
   the sum of terms
 proportional to the multi-particle rate
\begin{equation}
\mathcal{R}(p^2)\equiv \frac{1}{2M_h ^2 } \sum_n \int \mathrm{d}\Pi_n \left | \mathcal{M}(1\rightarrow n)\right|^2
\end{equation}
where $M_h$ is the Higgs mass, $\Pi_n$ is the $n$-particle phase space element and $\mathcal{M}(1\rightarrow n)$ is the matrix element for $1\rightarrow n$ Higgs decay. If one assumes that $\mathcal{R}(p^2)$ is exponentially growing in $p^2$, all predictive power for $\Delta_F$ from $\rho(p^2)$ is lost, due to the infinite number of subtractions  required for a convergent integral 
in Eq.\eqref{divFormTwo}. Although one may know the exact form of $\rho(p^2)$, one may add an arbitrary analytic function to both the left-hand side and right-hand side of Eq.\eqref{subForm} such that the Feynman propagator is allowed 
to  change  its functional form wildly without having any apparent effect on  the multi-particle rate $\mathcal{R}(p^2)$.

This feature is just a statement that for an order-N polynomial $g(z)$ with  a branch cut $\Delta g(z)$ along the real axis beginning at $z_0$, one can integrate  $\Delta g(z)$ via contour integration. In order to discard the contribution from the $|z|\rightarrow \infty$ curve, one performs $N$ subtractions such that 
\begin{equation}
g(z) = (z-y)^N \frac{1}{\pi}\int_{z_0}^\infty \frac{\Delta g(x)}{(x-z)(x-y)^N}\,\,\mathrm{d} x - (z-y)^N \, \frac{\mathrm{d}^{N-1}}{\mathrm{d}y^{N-1}}\left( \frac{ g(y)}{(y-z)}  \right)
\end{equation}
where the latter term is the residue at the $x=y$ pole \cite{Sugawara:1961}. The price one pays for convergence is the addition of an order-$N$ ``polynomial of integration'' which must be fixed by extra conditions of the theory.

Returning to the Higgspersion scenario, we would like to stress that given that $\Delta_F (p^2)$ may include an arbitrary analytic function of $p^2$
there is no reason why it should fall off exponentially with $p^2$ in the high-energy limit. In fact, Eq.\eqref{subForm} suggests precisely the opposite -- that the two-point function should grow uncontrollably in this limit. 
The discrepancy between the amplitude growth with $p^2$ we observe 
and the  exponential fall proposed in \cite{Khoze:2017tjt} arises because the latter was calculated using perturbation theory. Namely, the single-particle irreducible (1PI) Green's function $\Sigma(p^2)$ was summed into a geometric series in order to put $\Sigma$ into the denominator of $\Delta_F (p^2)$. However, if $\Sigma$  grows exponentially  with $p^2$, at sufficiently large $p^2$ this series is no longer convergent and one must instead use the form 
\begin{equation}
\Delta_F (p^2) = \frac{i}{p^2 - m_0 ^2} \sum_{n=1}^\infty \left( -i \Sigma(p^2)\, \frac{i}{p^2 - m_0^2}
\right)
\end{equation}
where $m_0$ is the bare mass of the theory. In this form $\Delta_F$ will uncontrollably grow with $p^2$, in agreement with Eq.\eqref{subForm}. In this way, the Higgspersion mechanism only compounds the unitarity violation in  the Higgsploding theory.

\section{Conclusions}

We have explored the Higgsplosion effect and related Higgspersion mechanism 
behind it in detail and have found its limitation and problems.

In particular, assuming the correctness of the Eq.\eqref{FKhoze} for $F(\lambda n,\epsilon)$
derived for $1\to n$ process in \cite{Khoze:2017ifq,Khoze:2018kkz}   beyond the thin-wall approximation, we have found that the amplitude for $1\to n$ process increases  exponentially 
rather than decreases at sufficiently high energies as stated in  \cite{Khoze:2017tjt}.
We have found this effect and the respective discrepancy because 
one cannot use the resummation of the  self-energy insertion when that self-energy
grows exponentially.   Since the respective   series is   divergent for sufficiently large momentum one can not re-sum it into a correction of the propagator. Previously\cite{Khoze:2017tjt} it was argued that such a correction will play a crucial role in ``shutting-off'' the propagator at sufficiently large energies and solving hierarchy problem. In the light of our finding we would like to state that such a resummation is not possible and that,
assuming Eq.\eqref{FKhoze} is correct,
 the $1\to n$ amplitude will  grow exponentially thereby  violating unitarity.

The fact that  Eq.\eqref{FKhoze} implies  unitarity violation leads us to   conclude
that this equation is likely not generic enough and that additional
higher order cross terms of  $O(\lambda^2n^2\epsilon)$ form in Eq.\eqref{FHG2} are expected to play an important role on restoration of unitarity. Indeed,  unitarity should be restored, since it was present in the theory in the first place from the hermiticity of the Hamiltonian. If some theory has a real  unitarity problem (which is, however,  not the case  of  the SM  framework we discuss here) one of the natural solutions could be a  composite nature of the Higgs boson which  at certain characteristic energy scales would  cure nonunitary growth via the respective form factor and the related new physics sector.

In the case of the Standard Model we conclude that the   $1\to n$  multi-scalar final state amplitude 
should be consistent with unitarity, but that in any case if it exponentially grows it can not be re-summed. Such behaviour
is not consistent with unitarity and does not provide a solution to the  hierarchy problem.
We believe that the correct evaluation of $1\to n$ amplitude for multi-scalar final states above the threshold requires an extension
of  Eq.\eqref{FKhoze} and remains still an open and very non-trivial problem.

\section*{Acknowledgments}

FB would like to thank the IPPP associateship.  The authors are
grateful to A. Pukhov, V. Khoze, A. Monin, V. Rubakov, D. Levkov for valuable
discussions.
AB acknowledges partial  support from the STFC grant ST/L000296/1.
AB also thanks the NExT Institute and   Royal Society Internationl Exchange grant IE150682,
partial support from the InvisiblesPlus RISE from the European
Union Horizon 2020 research and innovation programme under the Marie Sklodowska-Curie grant
agreement No 690575.

\providecommand{\href}[2]{#2}\begingroup\raggedright\endgroup

\end{document}